\pdfoutput=1
\pdfoutput=1
\pdfoutput=1
\pdfoutput=1
\pdfoutput=1
\documentclass[journal=jacsat,manuscript=article]{achemso} 
\usepackage{chemformula}  
\usepackage[T1]{fontenc}  
\usepackage{indentfirst} 
\author{Dushuo Feng}
\affiliation{Key Laboratory of Optical Field Manipulation of Zhejiang Province, Department of Physics, Zhejiang Sci-Tech University, Hangzhou 310018, China}
\author{Zhihao Guan}
\affiliation{Key Laboratory of Optical Field Manipulation of Zhejiang Province, Department of Physics, Zhejiang Sci-Tech University, Hangzhou 310018, China}
\author{Xiaoping Wu}
\affiliation{Key Laboratory of Optical Field Manipulation of Zhejiang Province, Department of Physics, Zhejiang Sci-Tech University, Hangzhou 310018, China}
\author{Yan Wu}
\affiliation{Key Laboratory of Optical Field Manipulation of Zhejiang Province, Department of Physics, Zhejiang Sci-Tech University, Hangzhou 310018, China}
\author{Changsheng Song}
\email{cssong@zstu.edu.cn}
\affiliation{Key Laboratory of Optical Field Manipulation of Zhejiang Province, Department of Physics, Zhejiang Sci-Tech University, Hangzhou 310018, China}
\alsoaffiliation{Longgang Institute of Zhejiang Sci-Tech University, Wenzhou, China}

\title{Classification of skyrmionic textures and extraction of Hamiltonian parameters via machine learning}


\begin{document}

\begin{abstract}
\noindent Classifying skyrmionic textures and extracting magnetic Hamiltonian parameters are fundamental and demanding endeavors within the field of two-dimensional (2D) spintronics. By using micromagnetic simulation and machine learning (ML) methods, we theoretically realize the recognition of nine skyrmionic textures and the mining of magnetic Hamiltonian parameters from massive spin texture images in 2D Heisenberg model. For textures classification, a deep neural network (DNN) trained according to transfer learning is proposed to distinguish nine different skyrmionic textures. For parameters extraction, based on the textures generated by different Heisenberg exchange stiffness ($J$), Dzyaloshinskii-Moriya strength ($D$), and anisotropy constant ($K$), we apply a multi-input single-output (MISO) deep learning model (handling with both images and parameters) and a support vector regression (SVR) model (dealing with Fourier features) to extract the parameters embedded in the spin textures. The models for classification and extraction both achieve great results with the accuracy of 98\% (DNN), 90\% (MISO) and 80\% (SVR). Importantly, via our ML methods, the skyrmionic textures with blurred phase boundaries can be effectively distinguished, and the concluded formation conditions of various skyrmionic textures, especially the skyrmion crystal, are consistent with previous reports. Besides, our models demonstrate the mapping relationship between spin texture images and magnetic parameters, which proves the feasibility of extracting microscopic mechanisms from experimental images and has guiding significance for the experiments of spintronics.
\end{abstract}

\section{Introduction}
\label{sec:I}

\indent Various spin textures observed from 2D magnets \cite{RN3556,RevModPhys.78.1} which reflect different phases are induced by the competition between magnetic anisotropy and short-range exchange such as Heisenberg exchange and Dzyaloshinskii-Moriya interaction (DMI) \cite{PhysRevLett.64.3179,PhysRevLett.69.3385,PhysRevB.69.214401,PhysRevB.71.224429,PhysRevLett.70.1006}. In many magnetic systems, theoretical and experimental studies have been carried out on the morphology and stability of rich magnetic domain configurations such as stripes, labyrinths, chevrons, and bubbles.\cite{PhysRevLett.78.2224,PhysRevB.59.3329,VEDMEDENKO1998391}. Meanwhile, the study of phase transition about chiral magnetic textures such as spin helices or chiral domain walls has devoted a lot of effort due to their long lifetimes \cite{RN1988}. Recently, topological chiral magnetic spin textures, such as skyrmions \cite{RN970,Nagaosa2013,RN2846}, merons \cite{PhysRevB.104.064432,wang2020polar}, and other toroidal topological states \cite{damodaran2017phase,PhysRevResearch.1.033109} have been hot topics because of their promising applications for future spintronic devices with high storage density and energy efficiency \cite{puebla2020spintronic,aar7043} due to their unique static and dynamic properties. In addition, skyrmions have been detected in many magnetic materials such as MnGe\cite{5b02653}, FeGe\cite{yu2011near}, CrTe$_2$\cite{feng2023strain}, CrISe\cite{0117597}, and 2D Janus materials\cite{094403}, which are theoretically verified that the asymmetric DMI\cite{RN130,PhysRev.120.91} plays the key role in generating and stabilizing such fascinating topological chiral textures. Although the spin configurations developed in magnetic systems are well known, the topological spin textures with spin chirality containing complex interactions are still not completely understood.

The properties of 2D magnets are often quantified through specific magnetic parameters, if the parameters of a given physical system are appropriately extracted, the characteristics of the system can be understood and predicted theoretically \cite{cartwright1984laws}. For example, the micro-magnetic model can evolve the spin configurations to the stable state just based on the magnetic Hamiltonian \cite{brown1964some}. The common magnetic Hamiltonian usually includes several terms. Specifically, Heisenberg exchange energy attempts to align neighboring spins; DMI favors the tilting of neighboring spins; The purpose of magnetocrystalline anisotropy energy is to align the spins with the anisotropy direction or perpendicular to the anisotropy axis\cite{wang2020machine}. These terms are determined by parameters as Heisenberg exchange stiffness ($J$), Dzyaloshinskii-Moriya strength ($D$), and anisotropy constant($K$), respectively \cite{Leliaert_2018}. In previous researches, great efforts have been devoted in magnets to extract their key parameters by using neutron scattering \cite{PhysRevB.14.4897,shirane1968spin}, Brillouin light scattering (BLS) or ferromagnetic resonance (FMR) \cite{PhysRevB.29.4439,eyrich2012exchange,PhysRevB.53.12166,klingler2014measurements}. However, the methods mentioned above are time-consuming and costly because of the reliance on the inevitable measurements of time-resolved dynamics\cite{wang2020machine,PhysRevLett.119.030402,PhysRevLett.113.080401}. Furthermore, with the latest development of Lorentz transmission electron microscopy and other magnetic observing technologies, the experimental images can provide more detailed information of spin configurations. Although the spin configurations are determined by the magnetic Hamiltonian, it is not an easy task to extract the exact values of Hamiltonian parameters only from images within the huge Hilbert space\cite{7572162}. Therefore, the extraction of Hamiltonian parameters from experimental or simulated images automatically and efficiently is urgently needed.

Machine learning (ML) algorithms have great abilities to learn from the labeled data and identify huge data sets, providing an effective and efficient approach in the study of various physical phenomena in condensed matter physics, from the representation of quantum states \cite{aag2302} to the discovery of phase transition \cite{rem2019identifying,PhysRevB.94.195105,van2017learning} and the recognition of conventional phases of matter\cite{carrasquilla2017machine}. On one hand, convolutional neural network (CNN) has been successfully applied to the following tasks: identifying the phases \cite{PhysRevB.98.174411}, detecting the phases from videos of spin lattices \cite{PhysRevApplied.16.014005}, and constructing low-temperature phase diagrams for models which includes anisotropy terms \cite{SALCEDOGALLO2020166482}. On the other hand, it was shown that CNN can be successfully used to predict magnetic features such as the chirality \cite{PhysRevB.99.174426},the topological charge \cite{PhysRevApplied.17.054022} and the DMI \cite{9387547} from 2D images of spin configurations. Most researches in this field only focus on the characterization of typical configurations rather than the intermediate and indistinguishable phases emerging from thermal fluctuations\cite{PhysRevB.105.214423}, and few works have been reported on the prediction of magnetic Hamiltonian parameters combined with spin texture recognition.

In this paper, we investigate the classification of skyrmionic textures and the extraction of magnetic Hamiltonian parameters in 2D Heisenberg model with Heisenberg exchange interaction ($J$), Dzyaloshinskii-Moriya interaction ($D$), and single-ion anisotropy ($K$) by ML and micromagnetic simulations (MS). We produce the datasets of 6000 spin textures containing nine different phases which are obtained from MS. For phase classification, we propose an automated labeling framework to divide the original images into several categories of spin phases via unsupervised ML algorithms as K-means and hierarchical clustering. With a cutting-edge algorithm Detection Transformer (DETR), we recognize skyrmions accurately and analyse the influence of magnetic field and temperature on the magnetic phase transition. A standard deep neural network (DNN) trained from transfer learning is applied efficiently for distinguishing phase boundaries among the topologically protected magnetic skyrmion states, spin spirals originating from the spin-orbit coupling and other mixed states with a high accuracy of 98\%. Then, we obtain accurate phase diagrams under the manipulation of temperature, magnetic field and magnetic Hamiltonian parameters $J$, $D$, $K$, and gets a remarkable agreement when comparing with the results from simulations. Moreover, we construct a multi-input single-output (MISO) deep learning model for the extraction of magnetic Hamiltonian parameters and it reaches a high accuracy of 90\% within a large parameter space. In addition, a support vector regression (SVR) model obtaining information from Fourier images is proposed for the extraction too, which reaches an 80\% accuracy.

\section{Methods}
\label{II}

\subsection{Spin Hamiltonian and Micromagnetic Simulation}
\label{A}

To obtain spin textures with different magnetic Hamiltonian parameters, we use the following spin Hamiltonian:
\begin{equation}
  \label{eq1}
H=\sum_{\langle i, j\rangle} J_{i j}\left(\vec{S}_{i} \cdot \vec{S}_{j}\right)+\sum_{\langle i, j\rangle} \vec{d}_{i j} \cdot\left(\vec{S}_{i} \times \vec{S}_{j}\right)+\sum_{i} K\left(S_{i}^{z}\right)^{2}+\sum_{i} \mu B S_{i}^{z}
\end{equation}

Here, $J_{i j}$ and $\vec{d}_{i j}$ represent the Heisenberg exchange coeﬀicient and DMI vector between spin $\vec{S}_{i}$ and $\vec{S}_{j}$, respectively. The third summation formula includes single ion anisotropy which represented by $K$, and $S_{i}^{z}$ represents the z component of $\vec{S}_{i}$. $B$ denotes the z-oriented magnetic field while $\mu$ being the magnetic moment of magnetic atom. In our simulations, we only take into account the nearest neighbor interactions. The isotropic exchange interaction is positive, which corresponds to the ferromagnetic case. The DMI has an in-plane orientation and is perpendicular to the corresponding inter-site radius vector. The MS\cite{RN109,RN2989} is performed by Spirit package \cite{PhysRevB.99.224414} with Landau-Lifshitz-Gilbert (LLG) equation\cite{LANDAU199251,1353448}. To avoid the nonuniversal effects of boundary conditions, we use a supercell on the 200 $\times$ 200 square lattice containing 40000 sites with periodic boundary conditions. 

\subsection{Textures Classification via DNN}
\label{B}

\begin{figure}[!htbp]
  \begin{center}
  \includegraphics[width=16.0cm]{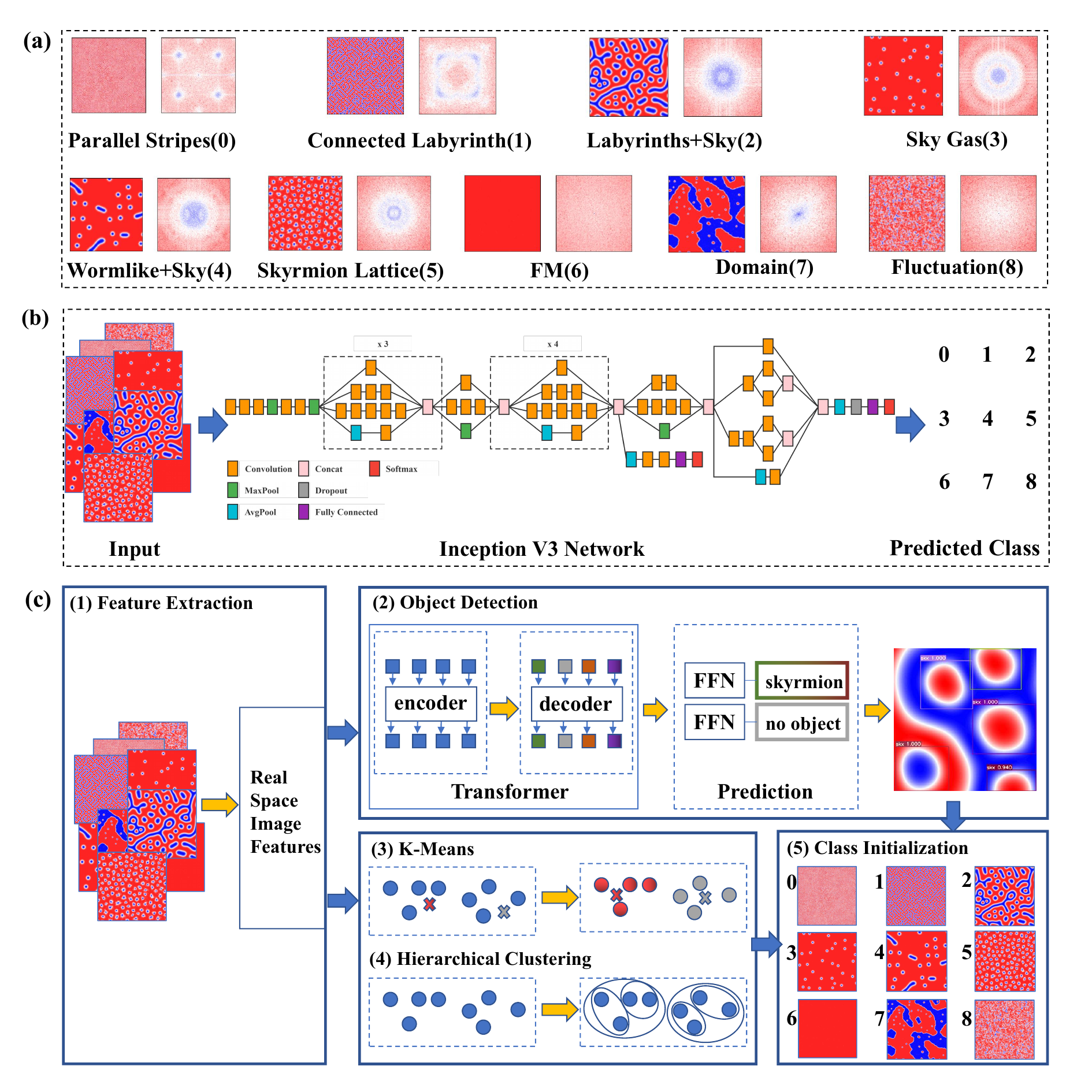}\
  \caption{\label{fig1}
  (a) The spin textures of nine phases obtained from micromagnetic simulation and their spectrum. (b) The architecture of the neural network used for phase classification. Simulated spin textures act as input data, entering the whole network and then match the corresponding classes. (c) Data preprocessing operation for automatically labeling and initializing dataset(shown as (5)) before training the deep learning model, including (1) feature extraction, (2) object detection algorithm, unsupervised learning methods such as (3) K-means and (4) hierarchical clustering. 
  }
  \end{center}
\end{figure}

As shown in Fig. \ref{fig1}(a), by using the Hamiltonian model presented in Eq. (\ref{eq1}) and MS, we produce datasets of spin textures for nine different spin phases [parallel stripes (0), connected labyrinth (1), labyrinths and skyrmion (2), skyrmion gas (3), fragmented ("wormlike") labyrinth domains and skyrmion (4), skyrmion lattice (5), ferromagnetic (6), domain (7) and fluctuation (8)] which contain classical spin configurations and unusual mixed phases. These noncollinear magnetic configurations obtained from Spirit simulations describing a 2D ferromagnet are projected on the z axis. Figure. \ref{fig1}(a) also includes spectrum diagram of each phase state, which analyzes the difference among each phase from the perspective of Fourier space. To ensure the adequacy and reliability of the datasets, we have simulated a total of 6000 spin texture images approximately with different configurations.

Then we employ a standard deep neural network architecture containing the fine-tuned Inception-v3 model based on transfer learning for textures classification. Figure. \ref{fig1}(b) presents an overview of our transfer learning experiment. The z-projected spin configurations belonging to different phases on the square lattice are considered as input data for the network. After training the model with these configurations, the output layer of the network is able to provide the recognized phases corresponding to the input images. For data preprocessing, since Inception-v3 model is based on CNN which is a supervised learning algorithm requiring labeled data, as shown in Fig. S1 (in Supplemental Material\cite{SM}), we propose an automated labeling framework to divide the original images into several different categories. The whole workflow of data preprocessing is shown in Fig. \ref{fig1}(c). We apply unsupervised learning algorithms such as K-means, hierarchical clustering (details see Fig. S2 of Supplemental Material\cite{SM}) for the initial division of categories. Specially, we apply object detection algorithm DETR (details see Fig. S3 of Supplemental Material\cite{SM}) \cite{978-3-030-58452-8_13} to identify phases containing skyrmions. Our approach streamlines the detection pipeline, effectively reduces the requirement for complex manual interventions. After data preprocessing, we may get almost accurate phases which can be trained in deep neural networks to achieve better performance of classification. 

\subsection{Parameters Extraction via MISO and SVR}
\label{C}

\begin{figure}[!htbp]
  \begin{center}
  \includegraphics[width=16.0cm]{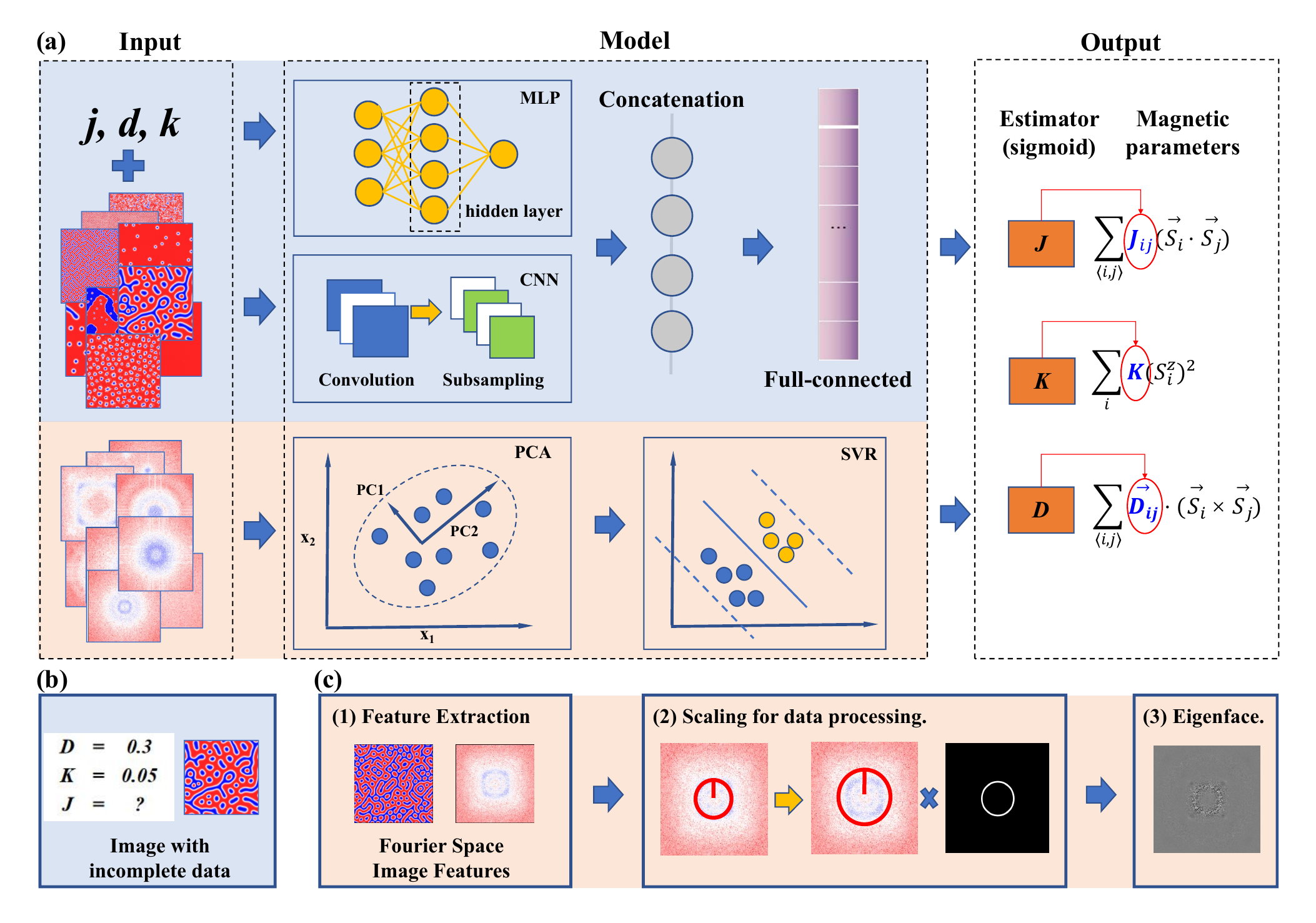}\
  \caption{\label{fig2}
  The architecture of the machine learning methods used for parameters determination. (a) Top row shows the MISO deep learning model which combines information from two sub-networks to detect the parameters. It takes two inputs: image features and numerical parameter features which are learned separately using sub-networks (CNN for images, MLP for numerical value). Bottom row shows the SVR which uses Fourier space features of images as input and outputs the estimated values of parameters. (b) The input data contains image data and the incomplete numerical parameter data for MISO model. (c) Data pretreatment for SVR regression model including (1) feature extraction from real space to Fourier space, (2) scaling images to filter out effective areas. After such pre-processing, eigenfaces which contain the most information of a single image are synthetized as shown in (3).
  }
  \end{center}
\end{figure}

Besides classifying different magnetic phases, we attempt to extract the physical microscopic parameters from spin texture images by ML. Two ML architectures are applied for magnetic parameters extraction. The first one defined as MISO model is shown at the top row of Fig. \ref{fig2}(a), which contains a CNN for images and a Multilayer Perceptron (MLP) for parameter labels. As shown in Fig. \ref{fig2}(b), it uses the spin texture images and the $J$, $D$, $K$ features of the images mentioned in section. \ref{B} as the multi-input variables, and aiming to estimate the magnetic parameters. This method allows us to accurately extract three key magnetic parameters ($J$, $D$, and $K$) from the input of simulated images and other fixed parameters. The second one is an SVR model (shown at the bottom row of Fig. \ref{fig2}(a)) using Fourier space images as input data and estimating the magnetic parameters. In order to extract features of the spin images to a greater extent, we take the Fourier transform of the original images and perform image scaling, then apply PCA to reduce dimensionality of the data and generates eigenfaces for SVR regression as displayed in Fig. \ref{fig2}(c). Such data preprocessing has been well applied in research on predicting DMI based on images\cite{PhysRevB.100.104430}, and details are shown in Fig. S13-S16 (in Supplemental Material\cite{SM}). These two methods mentioned above are used to extract magnetic parameters from different perspectives and processing modes.

\section{Results and discussion}

\begin{figure}[!htbp]
  \begin{center}
  \includegraphics[width=16.0cm]{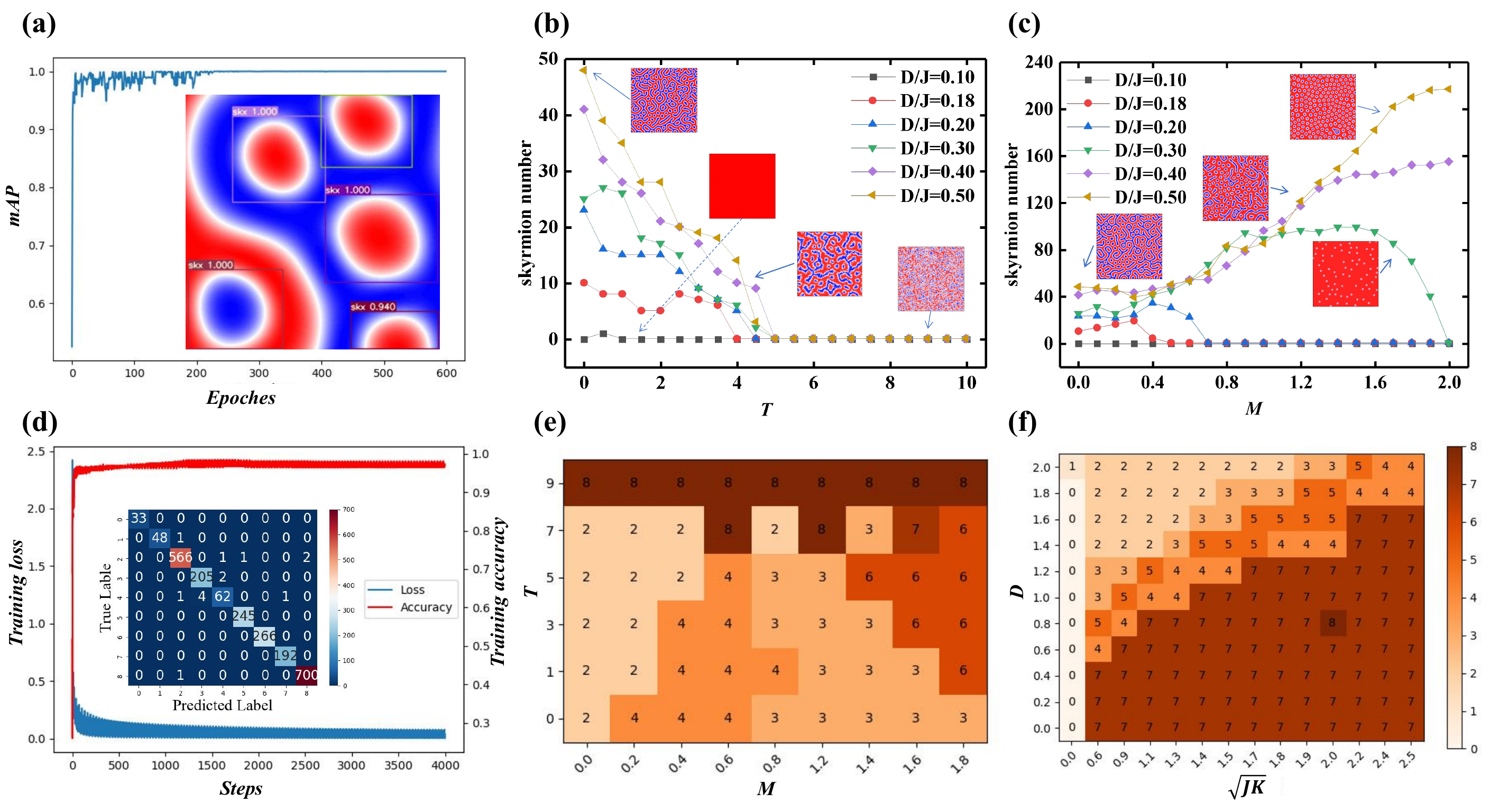}\
  \caption{\label{fig3}
  Training result and recognition result of ML methods used for phase classification. (a) The learning curves of mean Average Precision for DETR with the model trained for 300 epoches. Embedded illustration is the detection result of a texture containing five skyrmions. (b) The skyrmion number as function of temperature $T$ (unit is K). (c) The skyrmion number as function of magnetic ﬁeld $M$ (unit is T). (d) The change trend of loss and accuracy about the DNN model as the training steps increase. Embedded graph is the confusion matrices. (e) shows the M-T phase diagram at $D/J$=0.3 with K=0.25meV, the unit for $M$ and $T$ is T and K, respectively, and (f) shows the $\sqrt{J K}$-$D$ phase diagram at $M$=0.6T, the unit for $J$,$D$,$K$ is meV.
  }
  \end{center}
\end{figure}

The magnetic behavior dominated by the model presented in Eq. (\ref{eq1}) is influenced by magnetic Hamiltonian parameters, magnetic field and temperature. Based on such model, micromagnetic simulations allow us to obtain massive spin texture images for different configurations. As these simulations produce more variations and complex phases, we attempt to simply identify the amount of skyrmions and classify skyrmionic textures. Here we discuss the implementation of object detection algorithm to realize the recognition of skyrmion in detail. Figure. \ref{fig3}(a) gives the mean Average Precision (mAP, COCO style) learning curve of DETR, which is trained end-to-end with a set loss function (shown as Eq. (S1) in Supplemental Material\cite{SM}) and performs bipartite matching between predicted and ground-truth objects. To train the model, we label skyrmion as target to be identified based on the image data sets of Fig. \ref{fig1}(a), and reason about the relations of the skyrmions and the global image context to directly output the final set of predictions in parallel. We can see that the detected skyrmions are accurately located and plotted in the insert map of Fig. \ref{fig3}(a). When the learning rate decreases, DETR significantly boosts up the recognition performance, with a large mAP, which indicates this object detection model has achieved high accuracy.

Generally, different spin phases can be simply identified and distinguished by the number of skyrmions. After using DETR to recognize skyrmion accurately, we simultaneously investigate the evolution of spin phases under different ratio between DMI and Heisenberg exchange interaction ($D/J$), and take into account the influence of magnetic field ($M$) and temperature ($T$). Considering the influence of temperature, when $D/J$ ranges from 0.1 to 0.5 as shown in Fig. \ref{fig3}(b), the temperature takes the system from a stable phase containing skyrmions to a phase of thermal fluctuation (from mixed state 2 to state 8), which implies that the higher the temperature, the stronger the spin thermal fluctuation and the more unstable the skyrmion state. When $D/J$ falls inside the typical range around 0.2 for the formation of skyrmions, the skyrmions can be obtained at a relatively low temperature, which is consistent with the previous report\cite{fert2013skyrmions}. As shown in Fig. \ref{fig3}(c), when the magnetic ﬁeld increases, the system changes from a spiral phase (Sp, state 1) induced at zero ﬁeld to a skyrmion crystal (SkX, state 5), and then to a ferromagnetic (FM, state 6) state at higher $M$ for a small value of $D/J$. At a larger $D/J$,  the magnetic ﬁeld brings the system from a mixed phase containing skyrmions and labyrinths to a skyrmion crystal. These intermediate phases are enhanced by temperature and should disappear in the zero-temperature limit. When $D/J$ is relatively small, the exchange interaction plays a dominant role, which reduces the vortices in the FM background and makes the magnetic moments' rotation from the -z direction at the center of skyrmion to the z direction at the boundary at a small atomic scale. In this case, the skyrmions with relatively small-size and low-density can be obtained. Upon decreasing $D/J$ close to zero, DMI is negligible, and the strong magnetic exchange coupling $J$ leads the parallel arrangement of magnetic moments, that is, the FM phase. On the other hand, when $D/J$ is large enough, DMI plays a dominant role, benefiting for the emergence of labyrinth domains.

As the analysis above, we have discussed in detail the derivation of skyrmionic textures under the influence of Hamiltonian parameters, magnetic field, and temperature. By using nine spin phases generated under different parameters, magnetic fields and temperature, we train the standard DNN model for textures classification. The training accuracy and loss of this deep learning framework is shown in Fig. \ref{fig3}(d). The embedded illustration shows that all the test data are presented on the diagonal of the confusion matrix which describes the effectiveness of the classification for the test dataset, with predicted label as $x$ axle and true label as $y$ axle. The detailed phase diagrams taking into account the magnetic field (M) and temperature (T) as well as the ratio of the magnetic parameters can be seen in the following steps. As we can see in Fig. \ref{fig3}(e), when the parameters are set as $D/J$=0.3 and $K$=0.25meV, the phases containing skyrmions (state 2, 3 and 4) occupy the majority of the $M$-$T$ phase diagram. As magnetic ﬁeld increases, the system changes from the mixed phase of labyrinths and skyrmions (state 2) to the one of wormlike and skyrmion (state 4), skyrmion gas (state 3) and ferromagnetic (state 5) phase then appear at a higher $T$, which indicates that magnetic ﬁeld influences the stability of skyrmions. At the same time, the change in temperature affects the appearance of the entire phase diagram from the perspective of thermal stability. Based on this deep learning model, we also investigate the formation conditions of the skyrmion lattice. As we have established the roles of $D$ and $J$, we next take into account the effect of $K$ on the topological spin textures. As shown in Fig. \ref{fig3}(f), intriguingly, relationships between $D$ and $\sqrt{J K}$ of skyrmion lattice phase which referred to the state 5 are linear under magnetic fields of 0.6T (corresponding to the authoritative report\cite{2c00836}). This suggests that $D$/$\sqrt{J K}$ can be used to estimate the formation of skyrmion lattice, and further can be a judgment condition of the materials' properties.

\begin{figure}[!htbp]
  \begin{center}
  \includegraphics[width=16.0cm]{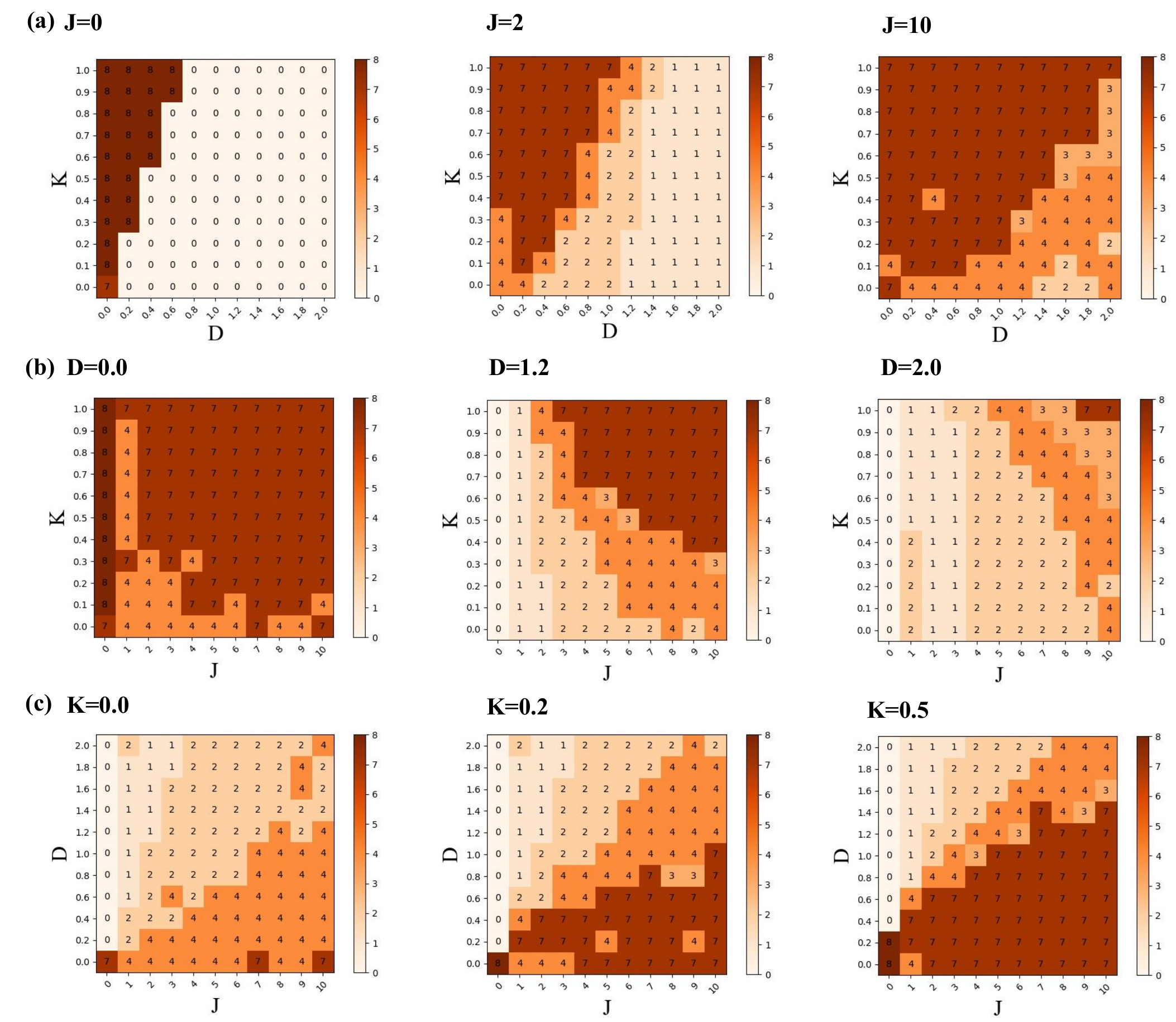}\
  \caption{\label{fig4}
  The phase diagram of $K$-$D$, $K$-$J$, $D$-$J$ are shown in (a), (b), (c) when $J$ set as 0, 2, 10 (meV); $D$ set as 0.0, 1.2, 2.0 (meV); $K$ set as 0.0, 0.2, 0.5 (meV); respectively. 
  }
  \end{center}
\end{figure}

As mentioned above, the $D$, $J$, and $K$ can significantly affect the formation and characters of different skyrmionic textures. Next, we attempt to identify a universal descriptor based on these three Hamiltonian magnetic parameters. As shown in Fig. \ref{fig4}, the value of $J$, $D$, $K$ can greatly determine a certain spin phase while adopting the control variate method, that is, each group of phase diagrams only studies one variable. As demonstrated in Fig. \ref{fig4}(a), we can see that the $J$ largely determines the skyrmion magnetic textures within the same regulatory regions of $D$ and $K$, in where $D$ ($K$) ranged from 0.0 to 2.0 (1.0) meV.  When $D$ increases, parallel stripes phase (state 0) or the connected labyrinth phase (state 1) turns to appear and expand at a low value of $J$. Then, the disconnected labyrinths (state 2) and skyrmion gas (state 3) comes with the higher $J$. The larger $J$ tends to break the disconnected labyrinth domain into fragmental domains mixed with skyrmions. In addtion, with the $K$ increases, domain phase (state 7) gradually dominant. With the $J$ increases, the distribution area of domain (state 7) become more, and overall, each mixed phase extends along the diagonal direction substantially. Similarly, when it comes to Fig. \ref{fig4}(b), with the increase of $D$, the parallel stripe phase (state 1) and the phase with few skyrmions (state 2-4) turn to expand. At the same time, domain phase (state 7) will be squeezed along the diagonal direction. Furthermore, if $D$ undergoes a larger value, the spin texture turns to contain the states of labyrinth domains (state 2) and parallel stripes (state 0) in a larger parameter space of $J$ and $K$. What can be concluded is that the increase of $D$ leads the density of labyrinth domains (state 2) to increase. What's more, with the $J$ increases, the fluctuation phases (state 8) may gradually dominant, and with the $D$ increases, the distribution area of domain phases (state 7) becomes less and comes with the appearance of skyrmion gas (state 3). After studying the phase distribution in phase diagram under the influence of $J$ and $D$, the impact of $K$ also follows the above patterns, which is shown in Fig. \ref{fig4}(c). As $K$ increases, the phases related to labyrinth domains (state 2) and skyrmions (state 4) occupy less area in the phase diagram. It can be understood according to the following description: when $K$ is positive, the energy gain provided by the perpendicular magnetic anisotropy turns to sustain an out-of-plane spin alignment, thereby promoting the formation of Néel-type skyrmions. In addition, similar to $J$, $K$ tends to force the magnetic moment in a parallel arrangement. In this regard, a large $|K|$ can suppress the modulation of labyrinth domains and skyrmions.

\begin{figure}[!htbp]
  \begin{center}
  \includegraphics[width=16.0cm]{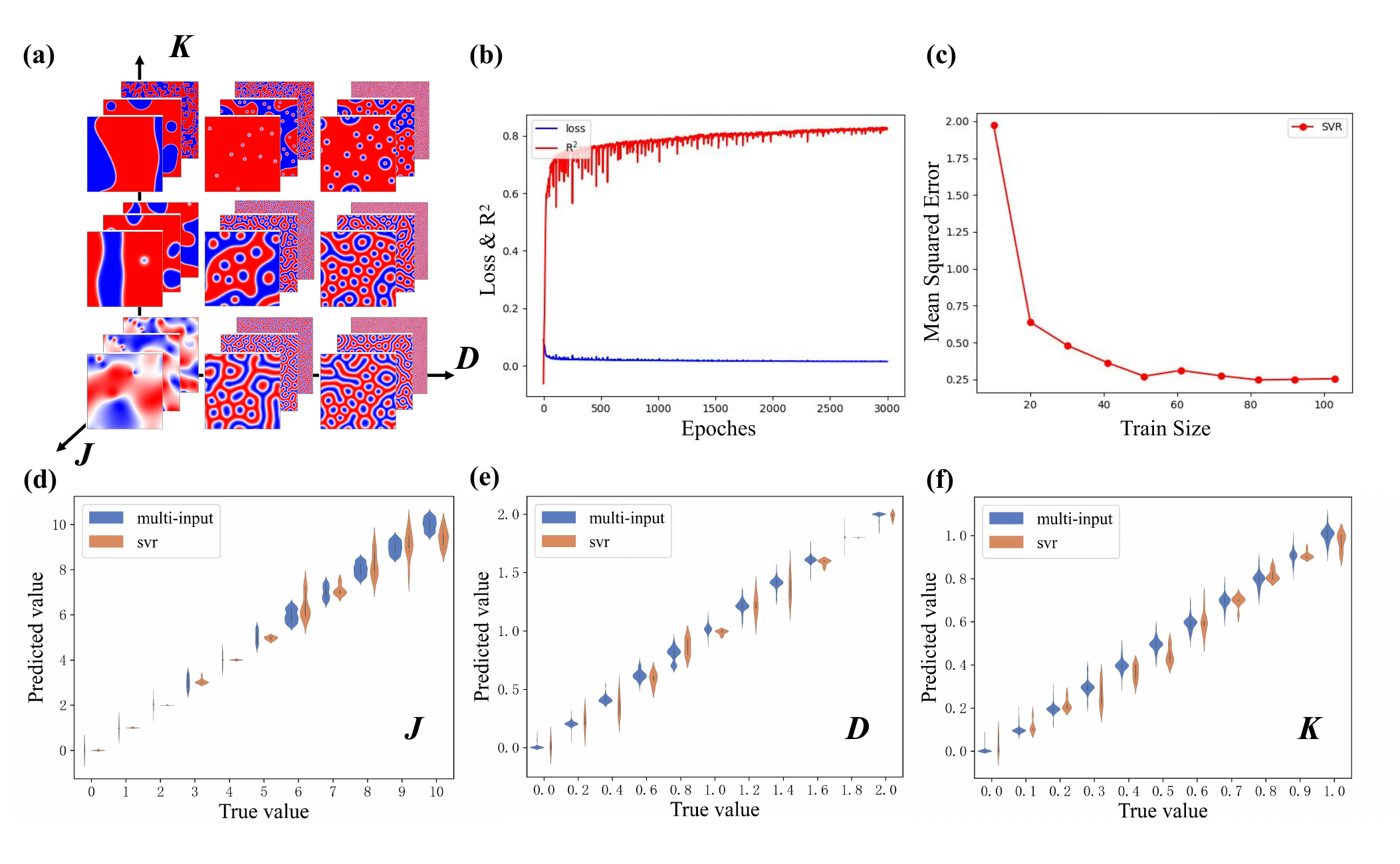}\
  \caption{\label{fig5}
   Regression prediction of Hamiltonian magnetic parameters tested on simulation images. (a) Spin configurations generated by MS with the parameters space of $J$, $D$, $K$ used in the training data. (b) The loss and $R^2$ for the parameters extraction of the MISO model. (c) The learning curve of the SVR model. (d)–(f) Plots of the estimated values as the function of true values of the $J$, $D$ and $K$, respectively.
  }
  \end{center}
  \end{figure}

Here comes to the ML models for extraction of Hamiltonian magnetic parameters. For a clear illustration, we show only the results of simulated images belonging to chiral ferromagnetic system ($J$ > 0). Figure. \ref{fig5}(a) shows a set of simulated spin configurations using a series of magnetic parameters $J$, $D$, $K$ as the training and testing datasets. These images of spin textures and their magnetic parameters of $J$, $D$, $K$ are input into MISO deep learning network mentioned in top row of Fig. \ref{fig2}(a), and the performance of the model is shown in Fig. \ref{fig5}(b), which shows the low loss and high $R^2$ meet good standards. Meanwhile, we perform Fourier transform on all images in the parameters space to obtain image features. Inputting the eigenfaces generated from the Fourier images via PCA into the SVR model introduced in bottom row of Fig. \ref{fig2}(a), we can get the same predicted sets of magnetic parameters $J$, $D$, $K$.  Figure. \ref{fig5}(c) gives the learning curve of SVR model, while the mean squared error (MSE) follows the calculation formula:$\operatorname{MSE}\left(y, y^{\prime}\right)=\frac{\sum_{i=1}^{n}\left(y_{i}-y_{i}^{\prime}\right)^{2}}{n}$. Here, $n$ is the number of samples, $y_{i}$ is the true value of the i-th sample, and $y_{i}^{\prime}$ is the predicted value of the model for the i-th sample. The learning curve shows that as the training size increases, the difference between the predicted and true values of the model decreases. In addition to proving the accuracy of the model through loss, accuracy and MSE curves, the matching results of the predicted values and true parameters can also prove the reliability of the model. The estimated values of $J$, $D$, and $K$ were plotted in Fig. \ref{fig5}(d)-\ref{fig5}(f), where the violin plots indicate the deviation between predicted results and the true values of the parameters. The values estimated by MISO model are marked in blue, while those estimated by SVR model are marked in yellow. It can be seen that there exist few errors for the extraction of these parameters through the two models. For example, the extraction of $J$ is fairly good when $J$ is not so large, and deviations of $D$ are slightly larger when it is in the middle range but still within an acceptable level. The predicted values and the true values follow a linear curve, indicating a generalization ability of extraction on all samples with different parameters sets. For comparison, the MISO model has higher accuracy in predicting $J$ and $D$, but is slightly inferior to SVR model in the process of predicting $K$.

\section{Conclusions}
\label{IV}

In this paper, we apply ML methods and MS to classify the skyrmionic textures and extract the magnetic Hamiltonian parameters in the 2D Heisenberg model with Heisenberg exchange interaction ($J$), Dzyaloshinskii-Moriya interaction ($D$) and single-ion anisotropy ($K$). From the images results of MS, we produce a large dataset with 6000 spin textures, and propose an automated labeling framework to divide the original images into nine different categories via K-means and hierarchical clustering. The DNN model based on inception-V3 architecture has achieved a high accuracy for the classification of skyrmionic textures and distinguishment of phase boundaries between the mixed phases. We recognize skyrmions accurately and simultaneously study the influence of magnetic field and temperature on phase transition with the cutting-edge algorithm DETR. Then, the accurate phase diagrams are obtained under the manipulation of the temperature, magnetic field and magnetic Hamiltonian parameters $J$, $D$, $K$, which gets a remarkable agreement when comparing with the results from simulations. The MISO model handling with both images and parameters and the SVR model obtaining information from Fourier images have achieved high accuracies of 90\%(MISO) and 80\%(SVR) for the extraction of Hamiltonian magnetic parameters. The precise models provided by our work can provide theoretical support for the classification of experimentally obtained skyrmionic textures and demonstrate the feasibility of extracting microscopic mechanisms of Hamiltonian parameters from skyrmionic images.

\begin{acknowledgement}

This work was supported by National Natural Science Foundation of China (Grant No. 11804301), the Natural Science Foundation of Zhejiang Province (Grant No. LY21A040008), and the Fundamental Research Funds of Zhejiang Sci-Tech University (Grant Nos. 2021Q043-Y and LGYJY2021015).

\end{acknowledgement}

\providecommand{\latin}[1]{#1}
\makeatletter
\providecommand{\doi}
  {\begingroup\let\do\@makeother\dospecials
  \catcode`\{=1 \catcode`\}=2 \doi@aux}
\providecommand{\doi@aux}[1]{\endgroup\texttt{#1}}
\makeatother
\providecommand*\mcitethebibliography{\thebibliography}
\csname @ifundefined\endcsname{endmcitethebibliography}  {\let\endmcitethebibliography\endthebibliography}{}


\end{document}